\documentclass[12pt]{article}
\usepackage{amsmath}
\usepackage{graphicx,psfrag,epsf}
\usepackage{enumerate}
\usepackage{natbib}
\usepackage{url} 
\usepackage{rotating,xcolor}

\newcommand{\BF}{\boldsymbol}

\begin{document}

\def\spacingset#1{\renewcommand{\baselinestretch}%
{#1}\small\normalsize} \spacingset{1}




\title{
    Variable Selection with Random Survival Forest and Bayesian Additive Regression Tree for Survival Data
	\vskip1em
}
\author{Satabdi Saha, Duchwan Ryu and Nader Ebrahimi}
\maketitle
\let\thefootnote\relax\footnotetext{
Satabdi Saha is Ph.D. student in Department of Statistics and Probability, Michigan State University, East Lansing, MI 48824 (Email: {\em sahasata@stt.msu.edu}).
Duchwan Ryu is Associate Professor, Department of Statistics and Actuarial Science, Northern Illinois University, Dekalb, IL 60115  (E-mail: {\em dryu@niu.edu}).
Nader Ebrahimi is Emeritus Professor, Department of Statistics and Actuarial Science, Northern Illinois University, Dekalb, IL 60115 (E-mail: {\em nebrahimi@niu.edu}).
}

\bigskip
\begin{abstract}
In this paper we utilize survival analysis methodology incorporating Bayesian additive regression trees to account for nonlinear and additive covariate effects. We compare the performance  of Bayesian additive regression trees, Cox proportional hazards and random survival forests models for censored survival data, using simulation studies and survival analysis for breast cancer with U.S. SEER database for the year 2005. In simulation studies, we compare the three models across varying sample sizes and censoring rates on the basis of bias and prediction accuracy. In survival analysis for breast cancer, we retrospectively analyze a subset of 1500 patients having invasive ductal carcinoma that is a common form of breast cancer mostly affecting older woman. Predictive potential of the three models are then compared using some widely used performance assessment measures in survival analysis literature.
\end{abstract}

\noindent%
\vfill

\newpage
\spacingset{1.45} 

\section{Introduction}


The identification of appropriate covariates and the adequate manipulation of non-linearity and high dimensionality are highly critical to investigate the effect of several important covariates on the event times and the accurate survival prediction.
%
The Cox proportional hazards (CPH) model seems to be the first natural step since it is the most commonly used in modeling event times with the covariates. It assumes a nonparametric form for the baseline hazard and allows testing for  differences in event times of two or more groups of interest, while allowing to adjust for covariates of interest. 
The CPH model, however, may result in biased parameter estimates if the covariates fail to follow the proportional hazards assumption. Moreover the assumption is often violated  due to the presence of complex relationships in the data structure. Failure of this assumption sometimes leads to the use of stratified Cox models that stratify with respect to covariates, or time dependent Cox models that incorporate a time dependent interaction term to deal with the non-proportionality of hazards. However several other problems including non-linearity, interactions between covariates and high dimensional parameter spaces are still hard to effectively address using CPH. Various nonparametric modeling methods such as penalized regression \citep{tibshirani1997lasso,zhang2007adaptive}, survival trees \citep{leblanc1993survival}, boosting with Cox-gradient descent \citep{ma2007clustering} and random survival forests \citep{ishwaran2008random} have been proposed to address the various problems faced by CPH model.

The use of classification and regression trees (CART) and other recursive partitioning methods has allowed for a more detailed study  of  the effects of covariates on the survival distribution. 
%
It has been proved to be an efficient tool in identifying significant covariates and their interactions. Its main advantages over other methods include effectively analyzing and interpreting complex nonlinear and high dimensional survival data and bringing a reduced dimension.  
However, CART method often suffers from high variation in prediction. Various methods that combine a set of tree models, so called ensemble methods, have attracted much attention to decrease the variance and to increase the prediction accuracy of CART.
These include boosting \citep{freund1995desicion,friedman2001greedy}, bagging \citep{breiman1996bagging} and random forests \citep{breiman2001random}, each of which uses different technique to fit a linear combination of trees.
%
These flexible nonparametric modeling methods have been successfully applied in the context of survival analysis. \cite{hothorn2004bagging} used the forms of bagging survival trees, \cite{hothorn2006survival} utilized survival ensembles,  and \cite{ishwaran2008random} considered random survival forests (RSF).

Some Bayesian models take the average of posteriors arising from single-tree models
as in \cite{chipman1998bayesian,Mallick1999,blanchard2004algorithme}. Such model averaging uses posterior probabilities as weights for averaging the predictions from individual trees. This idea has been further developed in a Bayesian sum-of-trees model, where each tree is constrained by a regularization prior to be a weak learner.
Motivated by ensemble methods, this sum-of-trees model known as Bayesian additive regression trees (BART) provides a framework to model the association of covariates and outcomes nonparametrically and flexibly.
BART models have shown an excellent predictive performance, for both continuous and binary outcomes, comparing to random forests and boosting \citep{chipman2010bart}. Recently the idea of BART has been extended to analyze survival data,
by expressing the nonparametric likelihood for the Kaplan-Meier estimator in a form suitable for BART \citep{sparapani2016nonparametric}. 
It has performed very well in terms of prediction error with medium variance and medium bias, while it has identified complex non-linear relationships and interactions in the dataset with efficient variable selection.


Among the proposed flexible methods, in this paper, we compare the performance of RSF and BART, as well as CPH, in the prediction of survival probability, in the variable selection and in the determination of marginal effects of the covariates. 
The models are then compared and contrasted  using simulation studies and a benchmark breast cancer dataset of 
invasive ductal carcinoma, sometimes called infiltrating ductal carcinoma, that is the most common type of breast cancer. According to the American Cancer Society, more than 180,000 women in the United States are diagnosed with invasive breast cancer each year and most of these are cases of invasive ductal carcinoma. Although invasive ductal carcinoma can affect women at any age, it is more common among women of older age. About two-thirds of women are 55 or older when they are diagnosed with an invasive breast cancer. There have been several studies reporting the effects of tumor size, tumor stage and tumor grade on the survival of breast cancer patients including \cite{d2001prognostic,rosenberg2005effect,delen2005predicting,omurlu2009comparisons,faradmal2014comparison}. We utilize the data extracted from the U.S. National Cancer Institutes' Surveillance, Epidemiology, and End Results (SEER) database for years 2005 to 2013 based on November 2015 submission that contains huge magnitude of data on several factors affecting breast cancer.

The remainder of this paper is as follows. In section 2 we review the three models chosen for analysis. Several performance assessment measures are reviewed to compare and contrast the predictive abilities of the models described. In section 3, we conduct simulation studies to compare the effectiveness of BART, CPH and RSF models in regression scenarios. In section 4, we use a real life dataset to demonstrate and compare the potential of the two ensemble learning methods. The survival experience of 1500 patients having invasive ductal carcinoma is studied using CPH, RSF and the BART models and their respective performances are compared using standard methods. 
Finally, conclusions of our findings are given in section 5.

\section{Methods}
\label{sec:meth}

For the $i$th subject, $i=1,\ldots,n$, consider $t_i$ as the actual event time and $c_i$ as the censoring time. Let $(z_i,\delta_i,\BF x_i)$ denote the survival data with the observed time $z_i=\min(t_i,c_i)$, the censoring indicator $\delta_i$ that takes 0 if the event is right censored and takes 1 otherwise, and the vector of  covariates $\BF x_i$.
In the analysis of survival time $T$ the intended purpose is to estimate the survival function $S(t)$ and hazard function $h(t)$ at time $t$ that are defined as
\begin{equation}
\begin{split}
	S(t)&= P(T>t)= 1 - F(t),\\
    h(t)&= \lim_{\Delta t \to 0} \frac{P(t<T  \leq t+\Delta t| T>t)}{\Delta t}=\frac{f(t)}{S(t)},
\end{split}
\end{equation}
where $F(t)$ denotes the cumulative distribution function and $f(t)$ denotes the probability density function at time $t$.
In the following subsections we briefly review three survival analysis methods: Cox proportional hazards, random survival forests, and Bayesian regression trees methods.
In addition, assessment of the prediction models and variable selection methods in survival analysis are discussed.

\subsection{Cox Proportional Hazards Regression}
\label{subsec: CPH}
The Cox proportional hazards model fits the survival data with the proportional hazard $h_{\textrm{Cox}}(t\vert\BF x_i)$ at time $t$ given the vector of covariates $\BF x_i$ such that
\begin{eqnarray}
 h_{\textrm{Cox}}(t|\BF x_i) = h_0(t)\exp(\BF\beta^T \BF x_i),					
\end{eqnarray}
where $\BF\beta$ is the vector of unknown regression coefficients and $h_0(t)$ is the arbitrary baseline hazard function. Hence, the cumulative hazard function $H_{\textrm{Cox}}(t\vert\BF x_i)$ and survival function $S_{\textrm{Cox}}(t\vert\BF x_i)$ are respectively given as 
\begin{eqnarray}
    \begin{aligned}
    H_{\textrm{Cox}}(t\vert\BF x_i) &= \int_{0}^{t} 
    h_{\textrm{Cox}}(v\vert\BF x_i) dv = H_0(t)\exp(\BF\beta^T\BF x_i),\\
    S_{\textrm{Cox}}(t\vert\BF x_i)&= \exp[\,-H_{\textrm{Cox}}(t\vert\BF x_i)],
\label{coxsurvival}	
    \end{aligned}
\end{eqnarray}
where $H_0(t)=\int_0^th_0(v)dv$ is the baseline cumulative hazard function.

\subsection{Random Survival Forests}
\label{subsec:RSF}

Random Forests are a machine learning ensemble method that combines the idea of bootstrap aggregation and random selection of features. 
%
%
%
For each of the bootstrap samples, it recursively splits the root nodes into interior and terminal nodes: (1) At each tree node a random selection of a subset of predictor variables is made; (2) Among all the binary splits made by the predictor variables selected in (1), the best split is determined using the predictor variable that maximizes the survival difference between daughter nodes; and (3) Repeat (1) and (2) recursively unless a terminal node has a minimum of $d_0$ $> 0 $ deaths. It calculates the cumulative hazard function (or survival function) for each tree and then averages over the $B$ bootstrap samples to obtain an ensemble cumulative hazard function (or ensemble survival function).


More specifically, for the observed time and the censoring indicator $(z_{il},\delta_{il})$ of the $i$th individual in the $l$th terminal node, consider  a censoring indicator $\delta_{il}$ that takes 0 for the event right censored and 1 for the event non-censored. 
Further, for the vector of covariates for the individual cases, consider the ordered distinct event times in the $l$th terminal node  $t_{il}$ as $ 0 < t_{(1l)} < t_{(2l)} <\cdots< t_{(kl)} < \infty$, where $t_{(jl)}$ denotes the $j$th ordered statistic among distinct observation and censoring times with $t_{(0l)}=0$. Denoting $d_{jl}$ and $R_{jl}$ as the number of deaths and the individuals at risk at time $t_{(jl)}$, respectively, the estimate for the cumulative hazard function (CHF) and the survival function with respect to are given as
\begin{eqnarray}
    \begin{aligned}
   H(t|\BF x_i)&= \sum_{t_{(jl)}\leq t} \frac{d_{jl}}{R_{jl}}, \text{\qquad for $\BF x_i \in l$},\\
    S(t|\BF x_i)&=\exp[\,-H(t|x_i)],\,\text{\qquad for $\BF x_i \in l$}.
    \end{aligned}
\label{eq:survRF}
\end{eqnarray}
The bootstrap ensemble CHF takes the average over the $B$ survival trees such that
\begin{eqnarray}
H_{\text{RF}}(t|\BF x_i)=\frac{1}{B}\sum_{b=1}^{B}H_b(t|\BF x_i),
\end{eqnarray}
where $H_b(t|\BF x_i)$ indicates the CHF obtained from a tree grown on the $b$th bootstrap sample. 


In a random forest, each of the trees is grown using an independent bootstrap sample from the set of training observations. 
One-third of observations are not used to construct a tree from the particular bootstrap sample. The remaining observations are referred to as out of bag (OOB) observations and are used to predict the ensemble CHF or ensemble survival function.
The resulting OOB predicted ensemble CHF or ensemble survival function are used as  valid test set predictions obtained from the random forest model. 

\subsection{Bayesian Additive Regression Trees}
\label{subsec:BART}

\cite{chipman2010bart} proposed Bayesian additive regression trees (BART) as a nonparametric Bayesian method that uses a Bayesian sum of trees model. BART enables full posterior inference including point and interval estimates of the unknown regression function.  Inspired from ensemble methods, each tree in the model is constrained by a regularization prior to be a weak learner, where fitting and inference are accomplished via an iterative Bayesian back-fitting algorithm.
For the $i$th individual let $y_i$ denote an outcome and $\BF x_i$ denote the vector of covariates. First, consider a single tree model and let $W$ denotes a binary tree consisting of interior and terminal nodes. A branch decision rule at each interior node typically splits the covariate space into two regions $\{\BF x_i \in S\}$ and $\{\BF x_i \notin S \}$, for a subset $S$ in the range of $\BF x_i$. Regarding a set of functional values $M = \{\mu_1,\mu_2,....,\mu_b\}$ associated with the $b$ terminal nodes and a function $g(\BF x_i,W,M)$ that assigns a $ \mu_i \in M $ to $\BF x_i$, the single tree model is described by 
\begin{equation}
    y_i= g(\BF x_i,W,M) + \epsilon_i,
\end{equation}
where $\epsilon_i$ is a Gaussian error with mean 0 and constant variance $\sigma^2$.
For $m$ trees, using $W$ and $M$ subscripted by $j=1,\ldots,m$, the BART model can be expressed as 
\begin{equation}
\label{bartmodel}
    \begin{split}
    y_i &= f(\BF x_i) + \epsilon_i,\\
    f(\BF x_i)&=\sum_{j=1}^{m} g(\BF x_i;W_j,M_j),
    \end{split}
\end{equation} 
where $\epsilon_i \sim N(0,\sigma^2)$.

For Bayesian specification of the sum of trees model, priors are carefully chosen to regularize the fit and to curtail strong individual tree effects such that

\begin{equation}
\label{eq:bartprior}
    \begin{split}
p[(W_1,M_1),\ldots,(W_m,M_m),\sigma^2]&=p(\sigma^2)\prod_{j=1}^m p(W_j,M_j)\\
 &=p(\sigma^2)\prod_{j=1}^m p(M_j| W_j) p(W_j)
    \end{split}
\end{equation} 
where $p(M_j|W_j)=\prod_{k=1}^{b_j} p(\mu_{jk}|W_j)$ for all $\mu_{jk} \in M_j$.
As \cite{chipman2010bart} noted, $p(W_j)$ has three aspects: (1) the probability of a node at depth $d$ to be non-terminal is calculated as $\alpha (1+d)^{-\zeta}$, where $\alpha \in (0,1)$ and $\zeta \in [\,0,\infty)$; (2) the choice of the splitting covariate at each interior node is uniformly distributed over the set of available covariates; and (3) the choice of a branching rule given a covariate at the interior node also follows an uniform distribution over the discrete set of available splitting values.
The default values, as specified by \cite{chipman2010bart}, are $\alpha=0.95$ and $\zeta=2$.
For $p(M_j|W_j)$, the conjugate normal prior on each of the value of the terminal node is available such that $\mu_{jk} \sim N(0,2.25/m)$. This prior has the effect of curbing strong individual effects of the trees so that every tree forms only a small part of the entire sum of trees model.
%
Regarding $p(\sigma^2)$, BART model for survival analysis uses a probit regression with standard normal latent variables and does not require a prior specification.



Applying BART for survival data, let $z_{(j)}$ denote the $j$th order statistic of distinct observed times such that $0<z_{(1)}<\cdots<z_{(k)}<\infty$ with $z_{(0)}=0$, $y_{ij}$  denote the event indicator for subject $i$ at $z_{(j)}$ and $p_{ij}$ denote the unconditional probability of an event occurring at time $z_{(j)}$.
As described by \cite{sparapani2016nonparametric}, the survival data $(z_i,\delta_i)$ are used to create the event indictors $y_{ij}$ that take $\delta_i$ when $z_i=z_{(j)}$ and 0 when $z_i<z_{(j)}$, for $j=1,\ldots,n_i$, where $n_i$ is the number of observation times less than or equal to $z_i$ such that $n_i=\mbox{number of }\{j:z_{(j)}\le z_i\}$. Then $\{y_{i1},\ldots,y_{in_i}\}$ will be either a sequence of $n_i$ zeros for a right censored event or a sequence of $n_i-1$ zeros and a one for a non-censored event.
%
%
With the BART function in \eqref{bartmodel} as a prior for $f$, the nonparametric probit regression model for $y_{ij}$ on time $t_{(j)}$ and the covariates $\BF x_i$ is given by
%

\begin{eqnarray}
    \begin{aligned}
    y_{ij}|p_{ij} &\sim \mbox{Bernoulli} (p_{ij}),\\
    p_{ij}|f &= \Phi(\,\mu_{ij}),\\
    \mu_{ij}&=\mu_0 + f(\,t_{(\,j)\,},\BF x_i),
    \end{aligned}
\end{eqnarray}
where $\mu_0$ is the mean of $f$ for centering.
For computational convenience, the truncated standard normal latent variables $l_{ij}$ are utilized such that 
\begin{align}
l_{ij}|y_{ij},f \sim \begin{cases}
	N(\mu_{ij},1)I(-\infty,0) & y_{ij}=0,\\
	N(\mu_{ij},1)I(0,\infty) & y_{ij}=1.\\
    \end{cases}
\end{align}
%
%
Through Bayesian back-fitting algorithm, Markov Chain Monte Carlo samples of all parameters and latent variables are available, and for any given covariate containing time $z_{(j)}$ and rest of covariates $\BF x_i$, Bayes estimator of $p_{ij}$, denoted by $p(z_{(j)},\BF x_i)$, is estimated by
\begin{equation}
     p(z_{(j)},\BF x_i)= \Phi[\mu_0 + f(z_{(j)},\BF x_i)].
\end{equation}
Further, the survival and hazard function at event or censoring time can be estimated as
\begin{eqnarray}
    \begin{aligned}
    S_{\text{BART}}(z_{(j)}|\BF x_i)&=P(T>z_{(j)}|\BF x_i),~~~j=1,\ldots,k,\\
    h_{\text{BART}}(z_{(j)}|\BF x_i)&= \frac{p(z_{(j)},\BF x_i)}{(z_{(j)}-z_{(j-1)})}.
    \end{aligned}
\end{eqnarray}
These functions can be only calculated at distinct survival times, however using the constant hazard assumption, interpolation between these times can be accomplished.


%

For exploring the effects of individual covariates and their specific effects on overall survival, the marginal survival functions involving single covariates or a subset of covariates can be calculated by using formulas derived in \cite{chipman2010bart}.
Consider a partition of covariates as $\BF x=(\BF x_{a},\BF x_{ib})$. Then, the partial dependence function of $\BF x_{a}$ can be defined as
\begin{eqnarray}
    f(\BF x_{a}) = \frac{1}{n}\sum_{i=1}^{n}f(\BF x_{a},\BF x_{ib})
\end{eqnarray}
and the survival function \citep{sparapani2016nonparametric} can be written as
\begin{eqnarray}
S(t|\BF x_{a}) = \frac{1}{n}\sum_{i=1}^{n}S(t|\BF x_{a},\BF x_{ib}).
\end{eqnarray}




\subsection{Performance assessment of the prediction models}
\label{method:perf}

An extremely important process in model building is assessing the prognostic competence of model. An important feature of this prognostic competence is discrimination, which refers to the ability of a predictive model to correctly classify subjects for their actual outcomes. In order to compare the discriminating potential of the aforementioned risk prediction models, we utilize some pre-existing methodologies such as concordance index (C-index), time-dependent Receiving Operating Characteristic (ROC) curves, Area under the ROC curve (AUC), Integrated area under the ROC curve (IAUC) and Integrated Brier Score (IBS) statistics.
In this paper, we consider time dependent ROC curves, IBS and C-index for the performance assessment.

ROC curves are very popular in displaying the sensitivity and specificity of a diagnostic marker say $\BF x_i$ and a disease variable say $D_i$ for subject $i$, $i=1,\ldots,n$. 
The disease status variable in survival analysis is often a time dependent variable.
%
Let $(z_i,\delta_i,\BF x_i)$ denote a survival data, where $z_i=\min(t_i,c_i)$ is the observed time for event time $t_i$ and censoring time $c_i$, $\delta_i$ is the censoring indicator that takes 0 for censored case and $\BF x_i$ forms the vector of risk set predictions. 
As \cite{heagerty2000time} proposed, let $D_i(t)=1$ if $t_i\leq t$ and $D_i(t) = 0 $ if $t_i>t$.
Then $D_i(t)=1$ indicates that the event has occurred prior to time $t$. Hence, at given time $t$ and a cut-off value $c$, the sensitivity and specificity are defined by
\begin{eqnarray}
    \begin{aligned}
    \mathrm{sensitivity}(c,t) &= Pr(\BF x_i>c|D_i(t)=1)= S_n(c,t),\\
    \mathrm{specificity}(c,t) &= Pr(\BF x_i\leq c|D_i(t)=0) = S_p(c,t).
    \end{aligned}
\end{eqnarray}
%
%
The ROC curve at time $t$ is defined as 
\begin{eqnarray}
ROC(c,t) = S_n[\{1-S_p\}^{-1}(\xi,t),t],
\end{eqnarray}
where $\{1-S_p\}^{-1}(\xi,t)=\inf\{c:1-S_p(c,t)\le \xi\}$.
This definition is often referred to as cumulative or dynamic ROC curve in literature because all the events that occur before time $t$ are considered as cases.
%
Other definitions of ROC curves can also be found in \cite{heagerty2005survival}.
The AUC statistic at time $t$ is defined as the area under the ROC curve at time $t$:
\begin{eqnarray}
    AUC(c,t)=\int ROC(c,t)dt
\end{eqnarray}

There are several available methods including Inverse Probability of Censoring Weights (IPCW) \citep{uno2007evaluating}, Conditional Kaplan-Meier  \citep{heagerty2000time}, Nearest Neighbor Estimator (NNE) \citep{heagerty2000time} and Recursive method  \citep{chambless2006estimation} for estimating the time dependent ROC curves.
In this paper we utilize the estimators of Uno based on IPCW, because it does not assume a specific working model for deriving the risk predictor. Furthermore, IPCW method assumes that censoring occurs independently of all the covariates, 
in addition to the non-informative censoring that the other methods assume.
%


The Brier score \citep{brier1950verification} is a quadratic score function that calculates the squared differences between actual binary outcomes $Y_i$, $i=1,\ldots,n$, and its predictions. In survival settings, let $Y_i(t)= I(z_i>t)$ denote the indicator of observed status of the subject $i$ and $\Hat{S}(t|\BF x_i)$ denote the predicted survival probability at time $t$ for subject $i$ with predictor variables $\BF x_i$. If an independent test data set with $\omega$ subjects $K_\omega$ is considered then the expected Brier score \citep{mogensen2012evaluating} is estimated by

\begin{eqnarray}
    BS(t,\Hat{S})
    =
    \frac{1}{\omega} \sum_{i \in K_\omega}
        \left[\frac{\left\{1-Y_i(t)\right\}\delta_i}{\Hat{G}(z_i\!\!- \vert \BF x_i)}+\frac{Y_i(t)}{\Hat{G}(t|\BF x_i)}\right] \left[Y_i(t)-\Hat{S}(t|\BF x_i )\right]^2
\end{eqnarray}
where $\Hat G$ is the Kaplan-Meier estimate of the censoring distribution and  $\Hat{S}$ is based on the training data.
IBS \citep{mogensen2012evaluating} can be calculated by integrating Brier score such that

\begin{eqnarray}
IBS(BS,\tau) = \frac{1}{\tau} \int_0^\tau BS(b,\Hat{S})db
\end{eqnarray}
where $\tau > 0$ can be set to any value smaller than the  maximum time of the test sample.
IBS ranges from 0 to 1; the
smaller the score, the better the fit, and serves as the most important benchmark when Kaplan-Meier estimate of survival is considered. 


As a global discrimination index, \cite{harrell1982evaluating,harrell1996tutorial} utilized the C-index to evaluates the predictive competence of a survival model that takes a value between 0 and 1, where 0 indicates the discordance and 1 indicates the concordance between observation and prediction. Among all possible pairs of patients ($i,j$) with observed times ($z_i, z_j$) and censoring indicator ($\delta_i, \delta_j$), consider the utilizable pairs that are not corresponding to $z_i < z_j$ when $\delta_i = 0$, $z_j < z_i$ when $\delta_j = 0$, and $z_i = z_j$ when $\delta_i = \delta_j = 0$.
If the predicted outcome is worse for the patient with the shorter observed survival time, the untied time utilizable pair is counted by 1 and the tied time utilizable pair is counted by 0.5. 
When both patients are censored, if the predicted outcomes are tied, the tied time utilizable pair is counted by 1, otherwise it is counted by 0.5.
When not  both patient are censored, if the censored patient has a worse predicted outcome, the tied time utilizable pair is counted by 1, otherwise it is counted by 0.5.
The C-index estimates the agreement probability between the observed and predicted survival outcomes by taking the ratio for the sum of all assigned counts of utilizable pairs to the total number of utilizable pairs.

\subsection{Variable selection in survival models}
\label{method:variables}

Among possible variable section methods, we review and utilize three methods that use Cox regression, random survival forest and BART, respectively. Firstly, using Cox regression within a backward stepwise method, the variables in each step are selected using the Akaike information criteria (AIC). The AIC criteria \citep{gneiting2007strictly} is closely related to the logarithmic scoring rule, which is strictly proper, and thus can be used for identifying a prediction model. This method can be implemented using the selectCox function in the R package pec \citep{mogensen2012evaluating}.

Secondly, using random survival forest, the OOB based C-index is obtained by dropping OOB cases down their in-bag survival tree and then assigning a daughter node randomly as soon as a split for the predictor variable is encountered. For each of the predictor variables, OOB prediction error is obtained by subtracting the OOB based C-index from one. A variable importance measure is then defined as the difference between the original OOB prediction error and the new OOB prediction error. Predictor variables having a large variable importance measure are considered to have greater prognostic capacities, whereas the variables with zero or negative variable importance measure can be dropped from the original model, as they add nothing to its predictive ability.

Lastly, using BART, the variables are selected by their appearance in the fitted sum of trees model. This method works better when the number of trees grown is small, as  growing a large number of trees can give rise to an inappropriate mix of relevant and irrelevant predictors, leading to redundancy  \citep{chipman2010bart}.   Variable selection is thus  accomplished by observing the individual predictor usage frequency in a sequence of MCMC samples as the number of trees grown becomes smaller and smaller. Thus predictors with a higher usage frequency  in the MCMC samples are considered to have higher prognostic competence as compared to the other predictors.
 

\section{Simulation Studies}

Semi-parametric methods such as the CPH model and the other parametric survival models  aims to model a particular functional relationship between the covariates and some survival outcomes. However BART  and RSF offers a more flexible approach allowing nonparametric functional relationships. In this subsection, we aim to study and compare BART, RSF and the CPH models via simulation models designed in \cite{sparapani2016nonparametric}

Two simulation studies are used, one having proportional hazards and the other having non proportional hazards. It is presumed that the model having non-proportional hazards should pose significant challenges to the semi parametric CPH model. Nine independent binary covariates $x=(x_1,x_2,x_3,x_4,x_5,x_6,x_7,x_8,x_9)$ are generated from the Bernoulli distribution with probability 0.5. They are then related to the Weibull event time t using survival function 
\begin{eqnarray}
    S(t)= \exp\left\{-\left(\frac{t}{\lambda}\right)^{\!\!\alpha} \right\}
\end{eqnarray}
with different rate and scale parameters for the proportional and non proportional hazards model. 

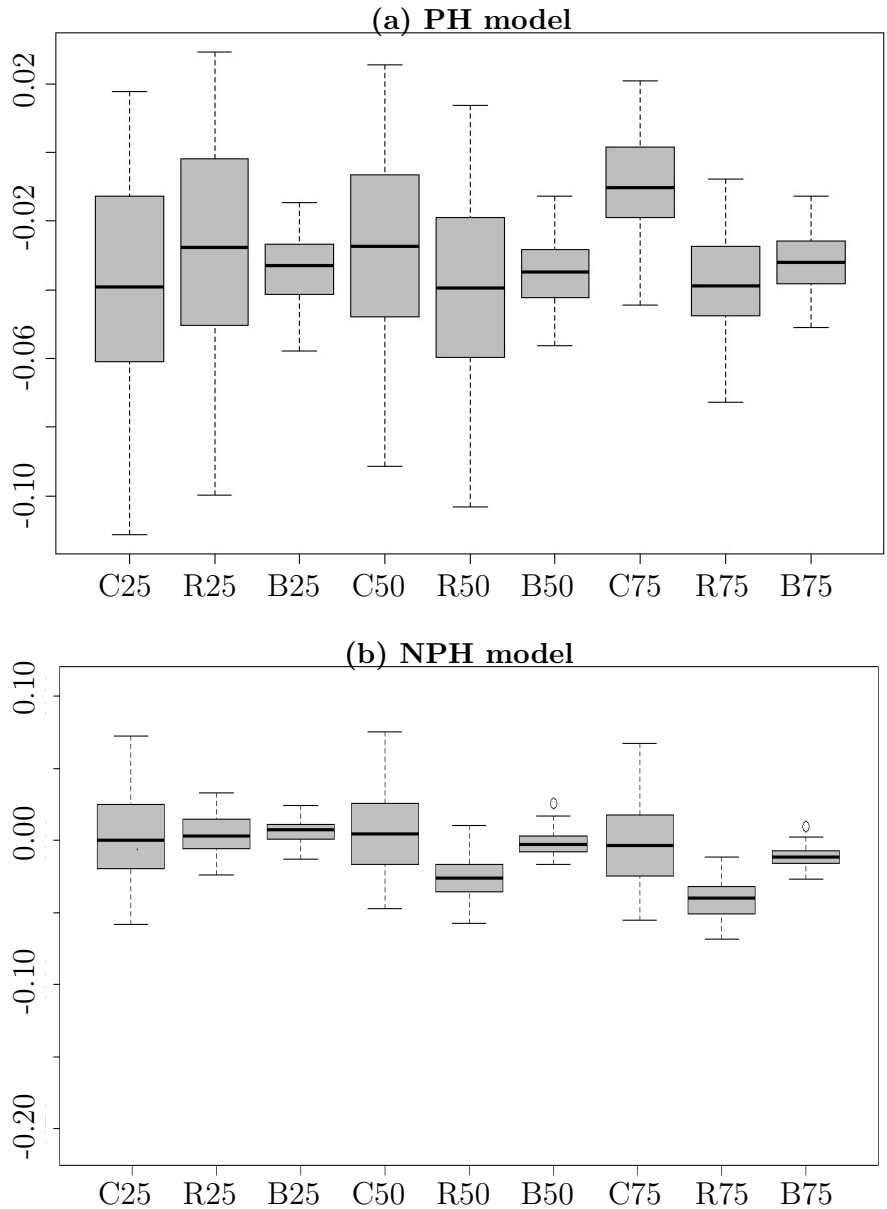
\begin{figure}[htb!]
\centering
$\begin{array}{c}
\includegraphics[height=2.8in,width=4.4in,trim=1.8cm 2.3cm 1cm 2cm,clip]{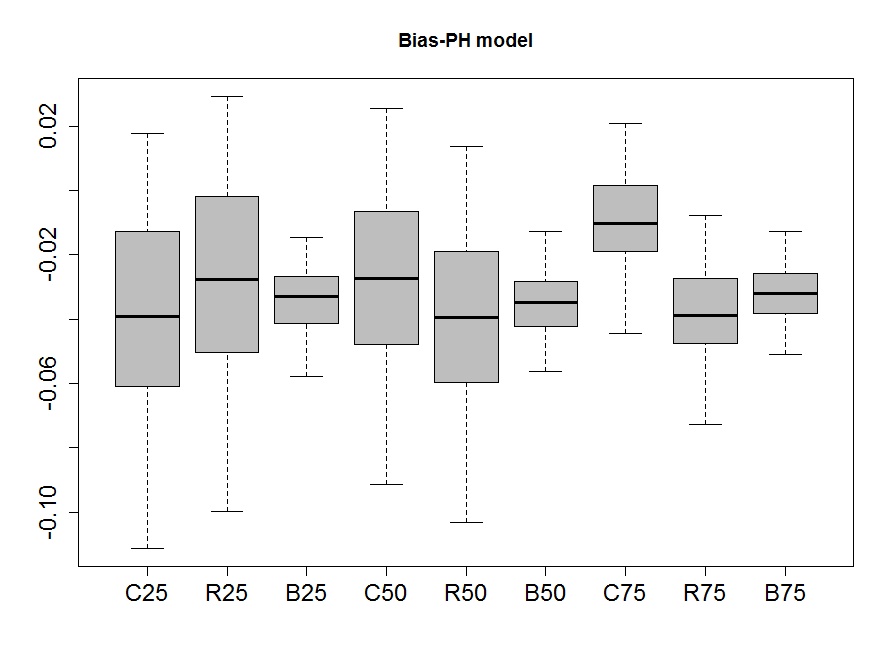}\\[2em]
\includegraphics[height=2.8in,width=4.4in,trim=2cm 3cm 1cm 2cm, clip]{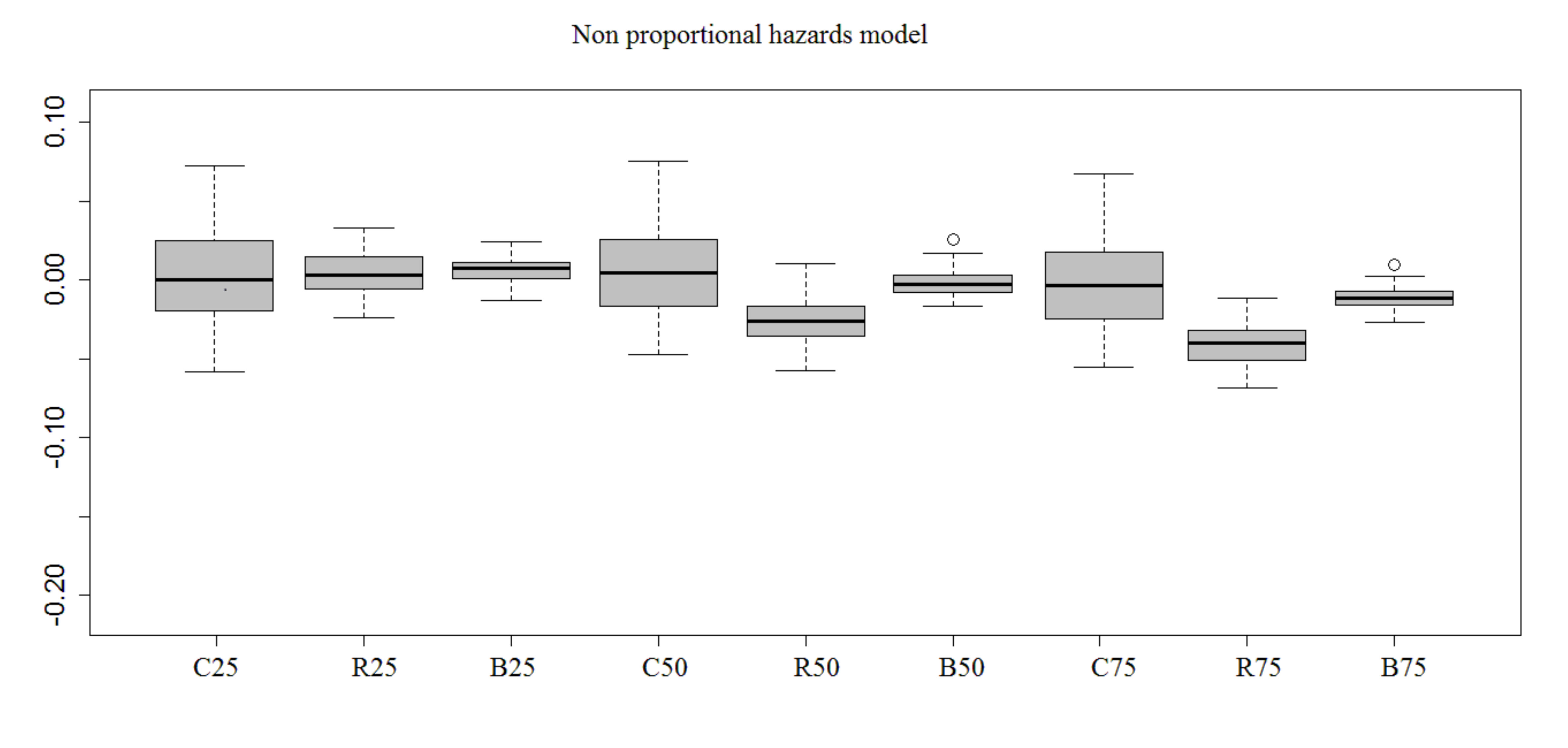}\\[1em]
\end{array}$
\put(-335,41){\begin{turn}{90}-0.10\end{turn}}
\put(-335,93){\begin{turn}{90}-0.06\end{turn}}
\put(-335,146){\begin{turn}{90}-0.02\end{turn}}
\put(-335,200){\begin{turn}{90}0.02\end{turn}}
\put(-303,15){C25}
\put(-271,15){R25}
\put(-239,15){B25}
\put(-207,15){C50}
\put(-175,15){R50}
\put(-143,15){B50}
\put(-110,15){C75}
\put(-77,15){R75}
\put(-45,15){B75}
\put(-335,-198){\begin{turn}{90}-0.20\end{turn}}
\put(-335,-143){\begin{turn}{90}-0.10\end{turn}}
\put(-335,-85){\begin{turn}{90}0.00\end{turn}}
\put(-335,-30){\begin{turn}{90}0.10\end{turn}}
\put(-303,-215){C25}
\put(-271,-215){R25}
\put(-239,-215){B25}
\put(-207,-215){C50}
\put(-175,-215){R50}
\put(-143,-215){B50}
\put(-110,-215){C75}
\put(-77,-215){R75}
\put(-45,-215){B75}
\put(-200,230){\textbf{\small(a) PH model}}
\put(-210,-10){\textbf{\small(b) NPH model}}
\caption{Box plots for bias}
\label{fig: BIAS plots}
\end{figure}


\begin{figure}[ht!]
\centering
$\begin{array}{c}
\includegraphics[height=2.8in,width=4.4in,trim=1.8cm 2.3cm 1cm 2cm,clip]{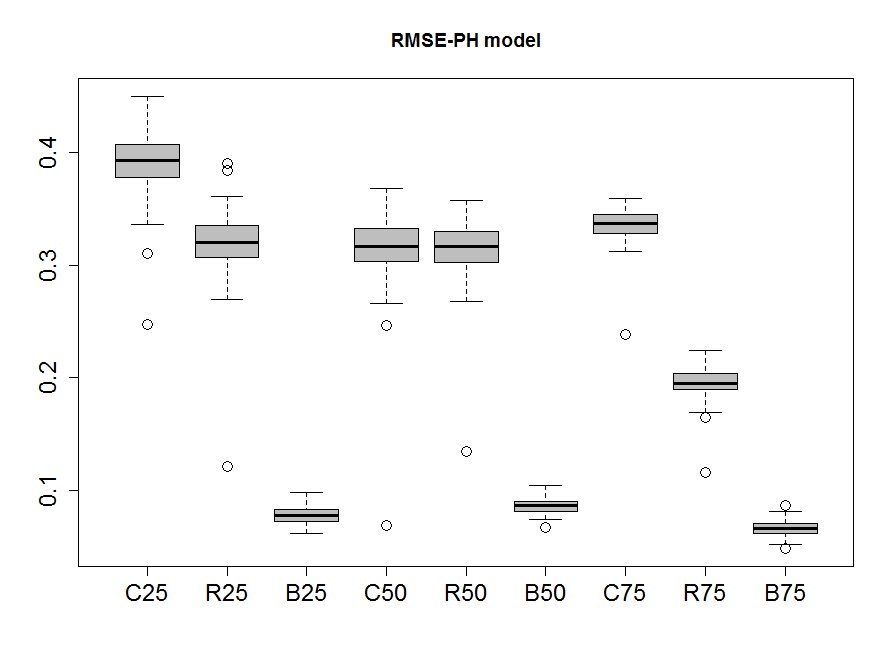}\\[2em]
\includegraphics[height=2.8in,width=4.4in,trim=2cm 3cm 1cm 2cm,clip]{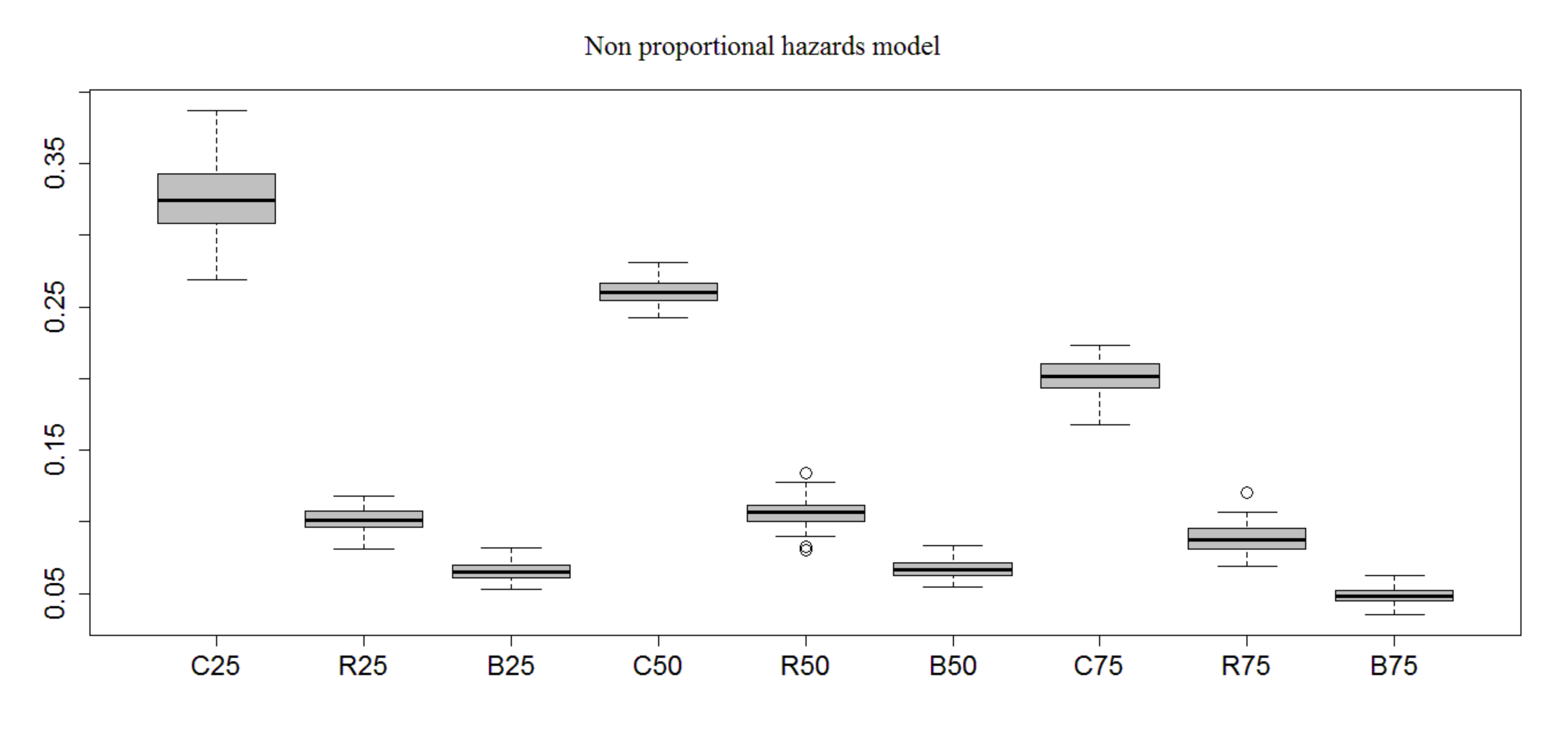}\\[1em]
\end{array}$
\put(-335,54){\begin{turn}{90}0.1\end{turn}}
\put(-335,100){\begin{turn}{90}0.2\end{turn}}
\put(-335,146){\begin{turn}{90}0.3\end{turn}}
\put(-335,191){\begin{turn}{90}0.4\end{turn}}
\put(-303,15){C25}
\put(-271,15){R25}
\put(-239,15){B25}
\put(-207,15){C50}
\put(-175,15){R50}
\put(-143,15){B50}
\put(-110,15){C75}
\put(-77,15){R75}
\put(-45,15){B75}
\put(-335,-193){\begin{turn}{90}0.05\end{turn}}
\put(-335,-143){\begin{turn}{90}0.15\end{turn}}
\put(-335,-94){\begin{turn}{90}0.25\end{turn}}
\put(-335,-45){\begin{turn}{90}0.35\end{turn}}
\put(-303,-215){C25}
\put(-271,-215){R25}
\put(-239,-215){B25}
\put(-207,-215){C50}
\put(-175,-215){R50}
\put(-143,-215){B50}
\put(-110,-215){C75}
\put(-77,-215){R75}
\put(-45,-215){B75}
\put(-200,230){\textbf{\small(a) PH model}}
\put(-210,-10){\textbf{\small(b) NPH model}}
\caption{Box plots for RMSE}
\label{fig: RMSE plots}
\end{figure}


For the proportional hazards model, we set 
\begin{eqnarray}
    \alpha =2,~\lambda = \exp\{3+0.1(x_1+x_2+x_3+x_4+x_5+x_6)+x_7\}.
\end{eqnarray} 
For the non proportional hazards model, we considered
\begin{eqnarray}
\alpha =0.7+1.3x_7,~\lambda = 20+ 5(x_1+x_2+x_3+x_4+x_5+x_6+10x_7).
\end{eqnarray}
Censoring times were generated independently from an exponential distribution with parameters selected to induce 20\% censoring. Samples of size 300 were considered and 100 datasets were generated, for each of the models. Each of the datasets were divided in  a 2:1 ratio for training and evaluation. Performances of the CPH, RSF and BART methods were compared based on measures of accuracy and bias, derived from the holdout test set of 100 samples, for both the model settings. The test root mean square error and bias are calculated for the CPH, RSF and BART survival prediction estimates. The expected value of the simulated estimates is then used as a measure of performance. Figure \ref{fig: BIAS plots} (a) and Figure \ref{fig: RMSE plots} (a) shows box plots of test set bias and RMSE for the proportional hazards model and Figure \ref{fig: BIAS plots} (b) and Figure \ref{fig: RMSE plots} (b) shows box plots of test set bias and RMSE for the non  proportional hazards model measured at the 25th, 50th and  75th percentiles of the overall survival distribution.
It is clear from these plots that the BART method performs closely to the CPH and RSF models in the proportional hazards case, however non parametric methods of BART and RSF performs significantly better than the semi parametric CPH model in the non-proportional hazards scenario.

\section{Application to breast cancer dataset}

In this section, we have tried to apply the CPH, RSF and BART models to a random sample of 1500 female patients between ages 24-90 having invasive ductal carcinoma as obtained from the U.S. SEER database for the year 2005.  Samples with missing data were not incorporated in the study, to facilitate the demonstration of methods. Ten covariates were considered in the analysis namely Age (in years), Race (White, Black, Others), disease stage (In-situ, localized, regional, or distant), tumor grade (well-differentiated, moderately differentiated, poorly differentiated, or undifferentiated), tumor size (in cms), estrogen receptor status (positive, negative or borderline), progesterone receptor status (positive, negative or borderline),  radiotherapy (received or denied), surgery (received or denied) and the number of lymph nodes. All the covariates selected for the analysis are important in breast cancer studies. Our response variable for the study was disease-specific survival (in months) based on the SEER cause-of-death code. Death from other causes was treated as censoring (non-informative censoring). The censoring times were assumed to be independent of the failure times. For evaluating the performance of the three models, the dataset was split randomly in the ratio 2:1 into a training set and a validation set.
Of the 1500 patient cases,  1081 patients were white, 298 black, and the rest were people from other different origins. A total of 1191 deaths occurred in the cohort of 1500  patients. The number of survival months (our outcome of study) ranges from 1-106 months. The mean follow up time was 34.75 months and the median follow-up months was 30 months. Most of the tumors were staged regionally (38.9\%). Most tumors were graded as poorly differentiated  (56.5\%). The mean age was 60.33 years with an SD of 14.59 years. The median tumor size was 29 mm with an IQR of  33 mm. 65.4\% of the tumors were estrogen positive. 43.7 \% of the patients received both surgery and radiation.

On application of the CPH model using the backward variable selection mechanism described in subsection \ref{method:variables},  race of the patient, age at diagnosis, tumor grade, tumor size, tumor stage, radiation therapy, surgery and estrogen receptor status are chosen as the most important variables. However, the accuracy with which the model estimates the  hazard ratios depends the proportional hazards assumption, which was violated by our dataset, thereby indicating the need for a more generalized model structure. We applied RSF models using log-rank splitting and log-rank score splitting, which ranked its covariates by level of OOB-importance, based on 1000 trees as described in subsection \ref{method:variables}. The five most important covariates in both the RSF approaches are surgery, tumor size, tumor stage, tumor grade and er status with a slightly different ranking. The bottom five covariates based on importance values are ranked similarly for both the RSF models. The top five predictors having maximum variable importance values were also selected by the Cox model. However predictors race, age at diagnosis and radiation therapy,  selected by the Cox model, have very low variable importance values for both the RSF models and are therefore considered unimportant for prediction purposes. 

\begin{figure}[ht!]
    \centering
    $\begin{array}{cc}
        \includegraphics[width=0.45\textwidth,trim=2mm 0 12mm 0,clip]{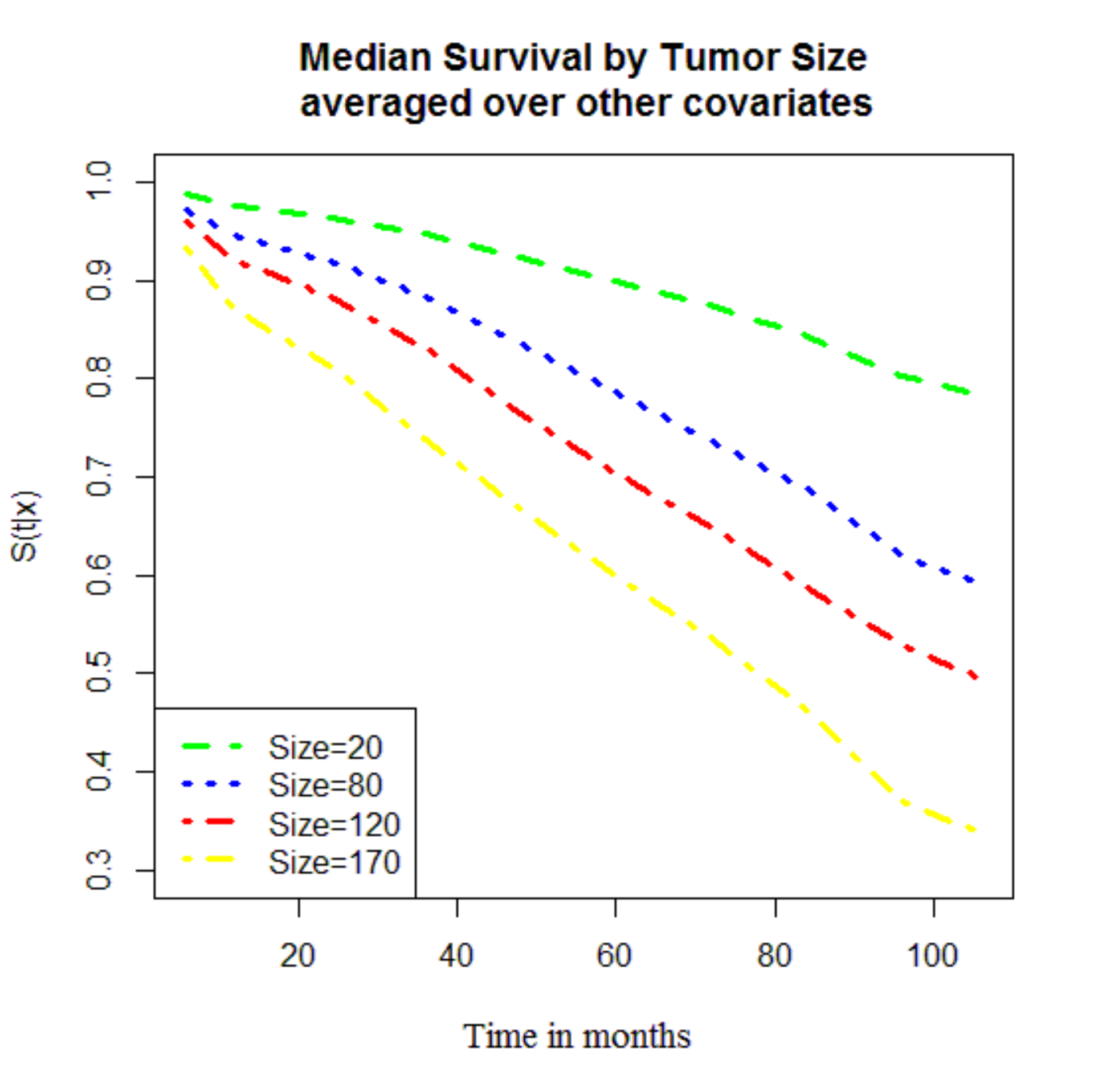} 
        &\includegraphics[width=0.45\textwidth,trim=2mm 0 12mm 0,clip]{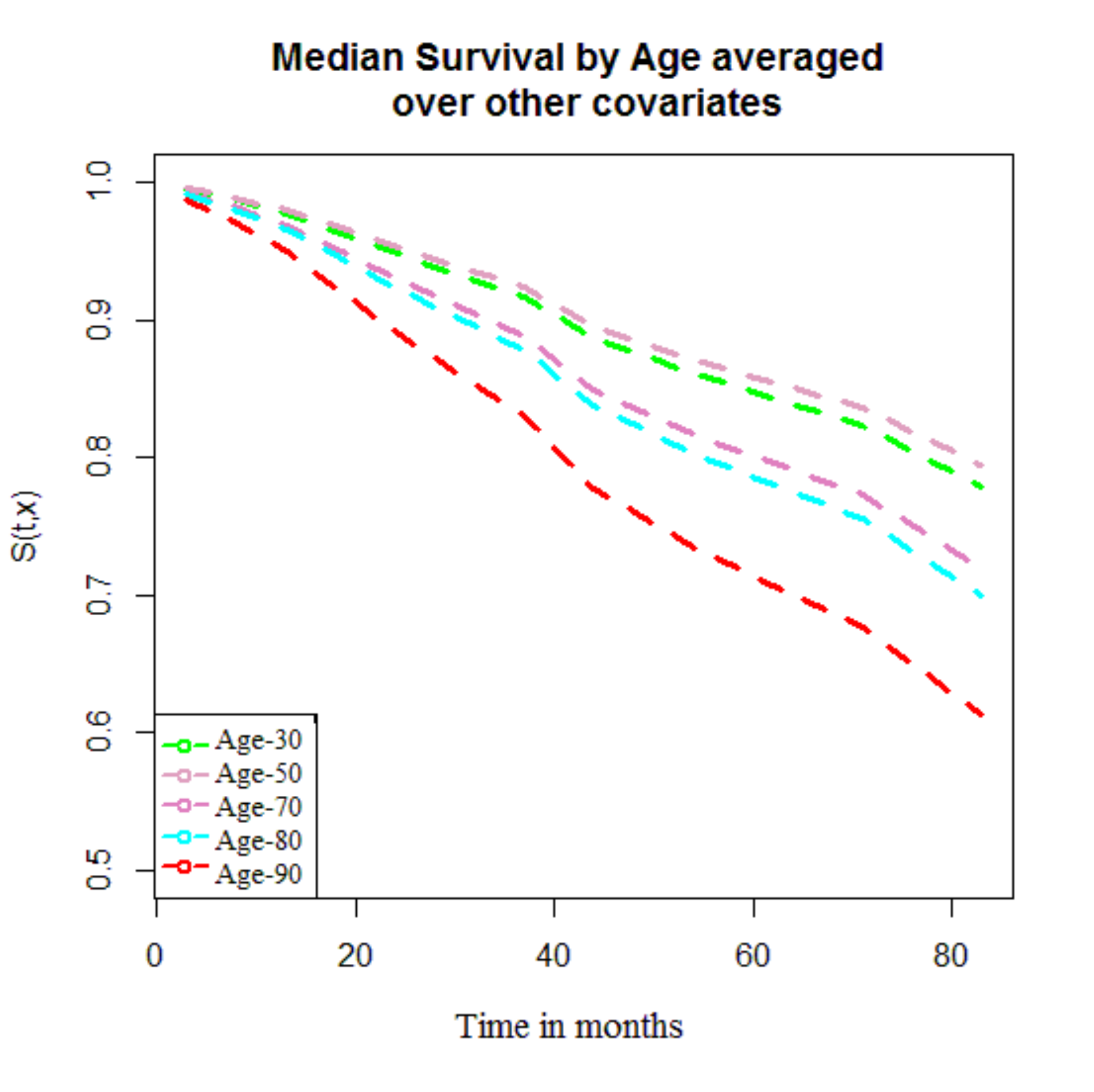} 
    \end{array}$
    \caption{{Marginal median survival functions.} The left plot is for different tumor sizes and the right plot is for different ages.}
    \label{Marginal Bart plots}
\end{figure}

The BART survival model was fit to the training dataset  with 50 trees in the sum and the default prior, having a burn-in  of 5000 draws and a long chain of 10,000 draws from the posterior distribution after the burn-in, for estimating the survival function given the predictors. For convergence checks, we generated several chains with different initial values and found comparable results. We obtained the partial dependence survival functions using equation for a particular subcategory of predictors. These functions can be explained as a marginal survival function for a single predictor level, averaged across the distribution of the remaining predictors. In Figure \ref{Marginal Bart plots} we have plotted the partial dependence functions for four different tumor sizes and five different ages. From the right plot in Figure \ref{Marginal Bart plots} it can be seen that survival probability drops rapidly with an increase in tumor size. For example the five year survival probability for a patient with tumor size 20 mm is 0.93 as compared to 0.71 for a patient with tumor size 120 mm. From the left plot of Figure \ref{Marginal Bart plots} it can be seen that the 5 year survival probability for a 50 year old breast cancer patient is 0.87 compared to 0.80 for a 70 year old patient.

\begin{figure}[ht!]
    \centering
    $\begin{array}{cc}
        \includegraphics[width=0.45\textwidth,trim=2mm 0 12mm 0,clip]{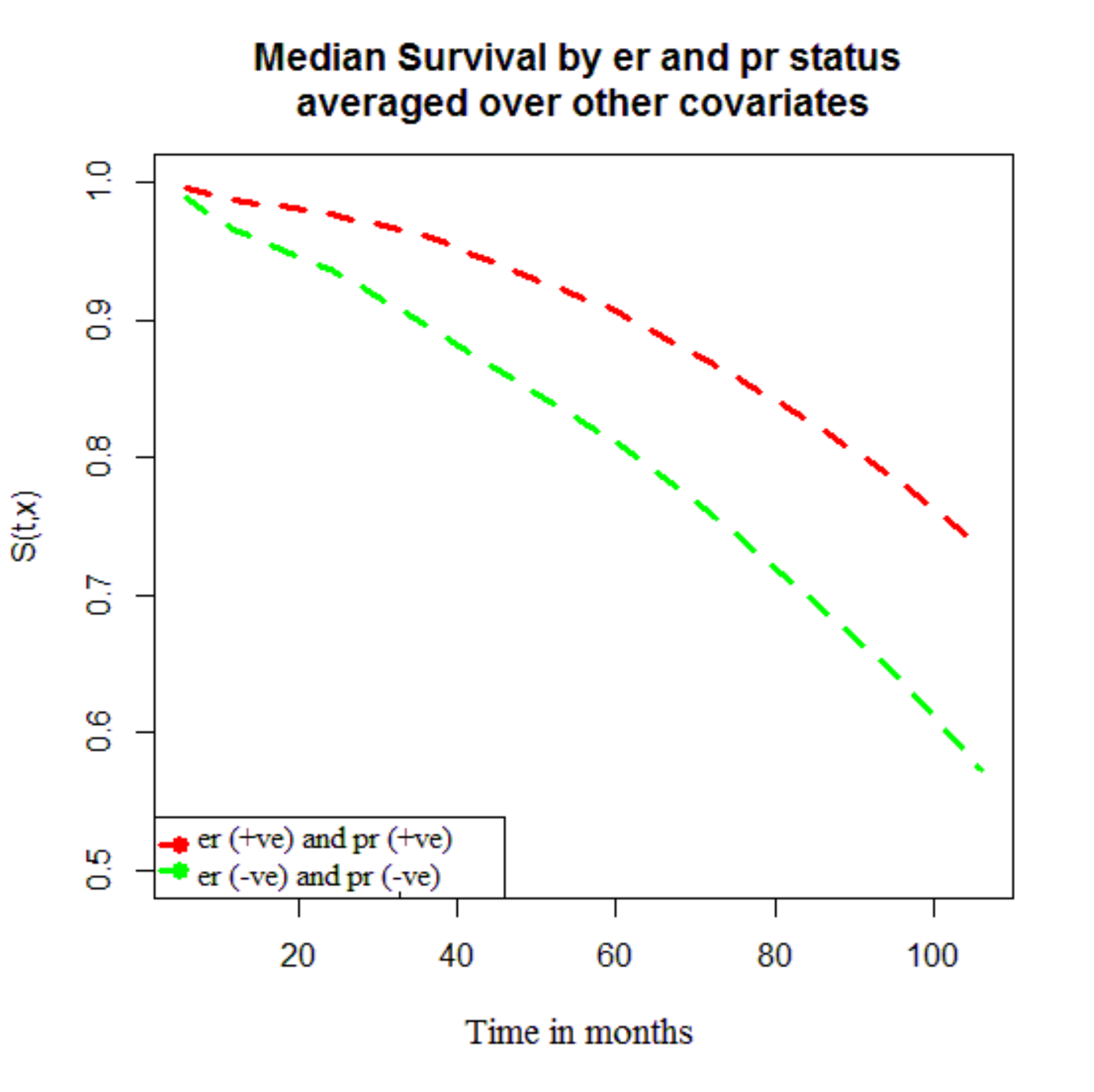} 
        &\includegraphics[width=0.45\textwidth,trim=2mm 0 12mm 0,clip]{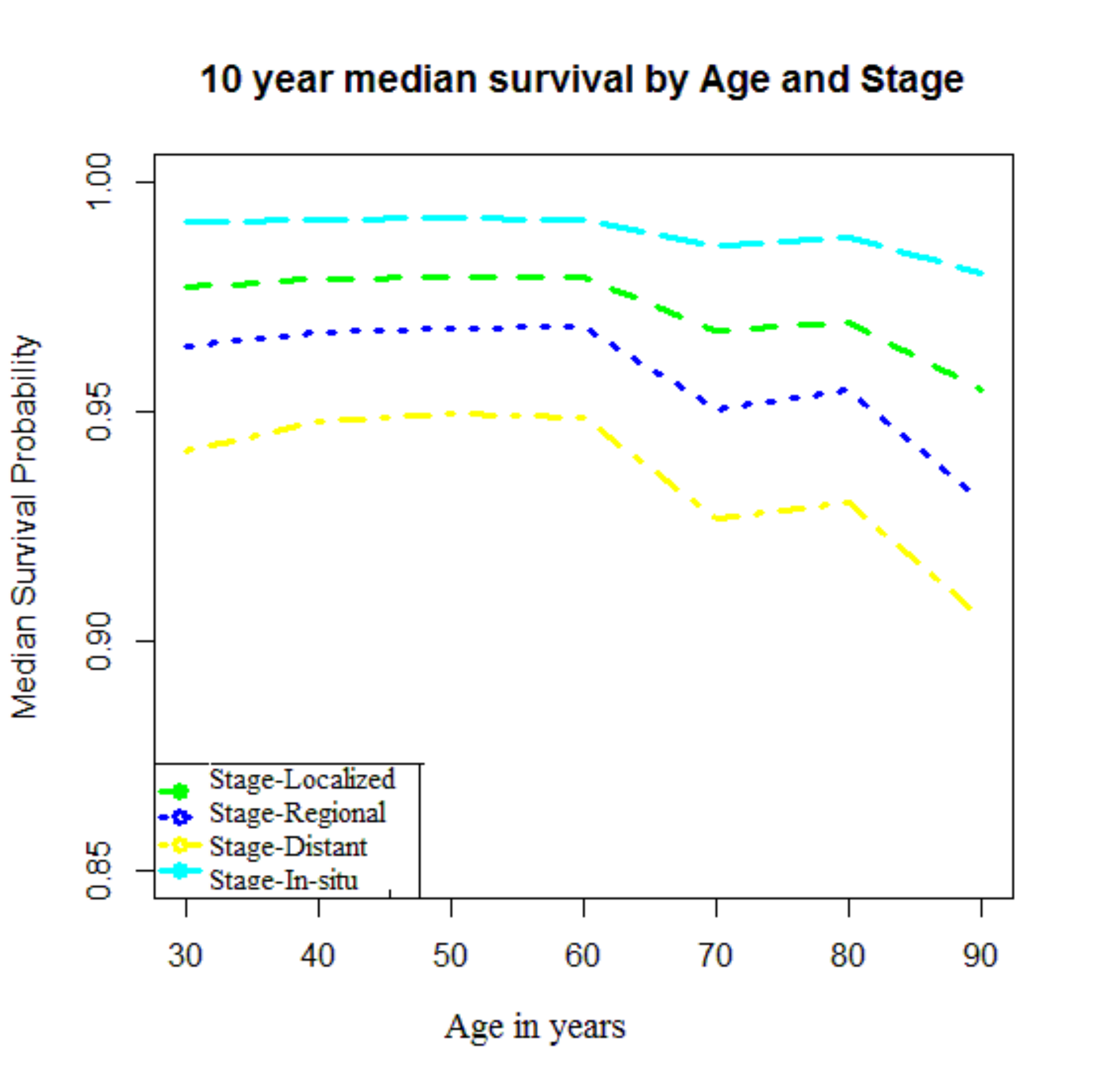} 
    \end{array}$
    \label{Median Survival}
    \caption{{Median survival functions}. The left plot is for different er and pr statuses and the right plot is for different stages.}
\end{figure}

The BART survival model can also be used to study the effect of interactions between covariates on the survival outcome. The left plot in Figure \ref{Median Survival} studies the effect of the interaction between estrogen and progesterone receptor status on the survival probability. It can be seen from the plot that estrogen receptor (er) and progesterone receptor (pr) positive breast cancer patients have a higher survival probability as compared to er and pr negative patients. Since Age and Stage variables were selected by both the CPH and RSF models we wanted to check whether there exists any interaction between them. There is no evidence of interaction as the right plot in Figure \ref{Median Survival}  shows nearly parallel patterns, while there may be an indication of a nonlinear relationship between median survival probability especially at age 70.

\begin{figure}[ht!]
\centering
\includegraphics[scale=.37, trim = .1cm 0cm 6cm 0cm, clip] {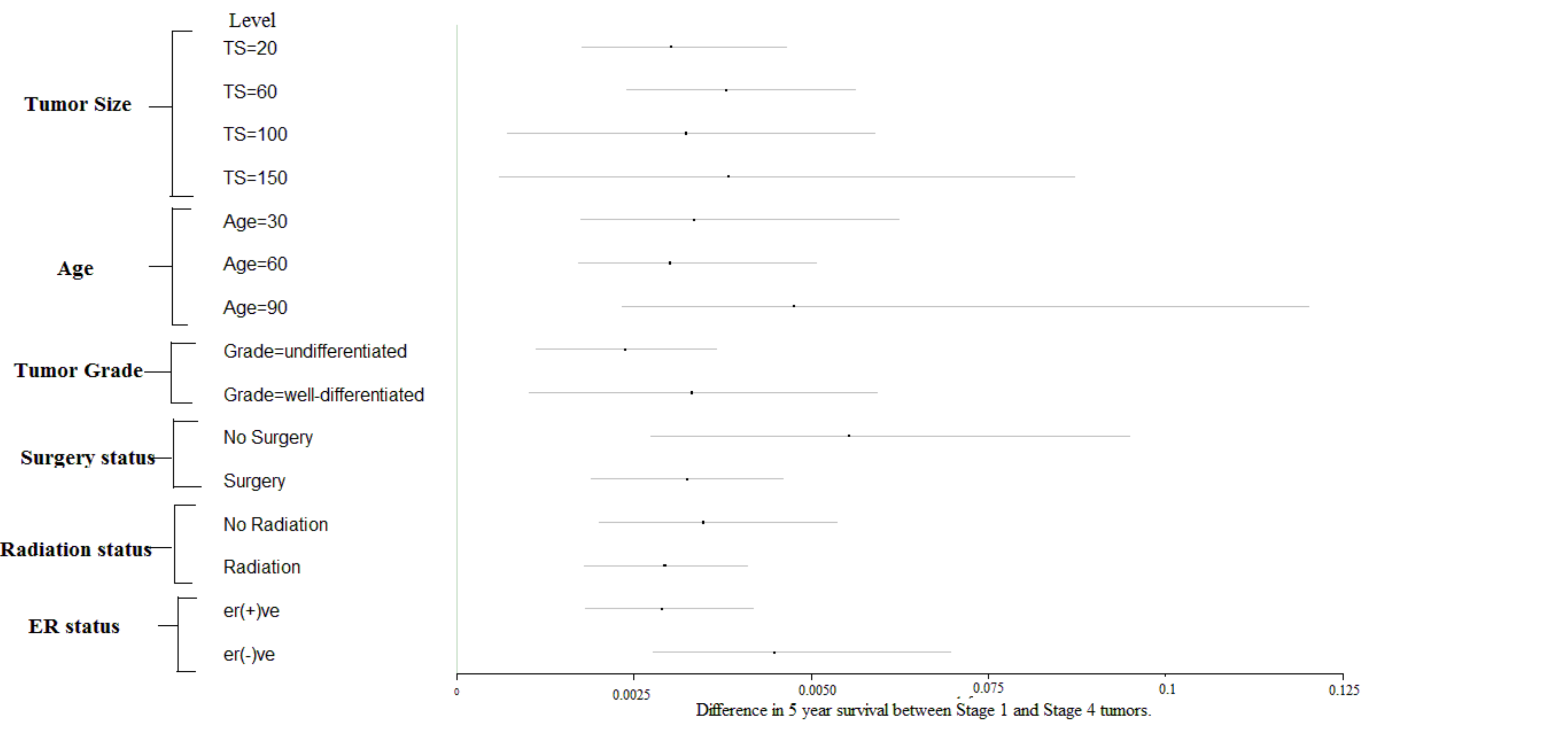}
\caption{Forest plot of the difference in 5 year survival between Stage 1 and Stage 4 by several covariates.}
\label{BART Forest Plot}
\end{figure}

BART can also be conveniently used, to draw inference on various aspects of the survival distribution (obtained by regressing on all or a subset of covariates), directly from the posterior samples. As another illustration on exploring significant interactions between the covariates, we explored the difference in the partial dependence survival function at five years  between patients having Stage 1 tumor and Stage 4 tumor, separately by tumor size, age, tumor grade, surgery status, radiation status and estrogen receptor status. These variables were selected as they have been considered important by both the CPH and the RSF models. Results obtained are shown as a forest plot in Figure \ref{BART Forest Plot}. One of the results indicated by the plot would be that surgery decreases the 5-year survival across the disease stages,  although the magnitude of the effect may vary slightly. 

\begin{figure}[ht!]
\centering
\includegraphics[scale=.34, trim=1cm 0 0.5cm 0, clip]{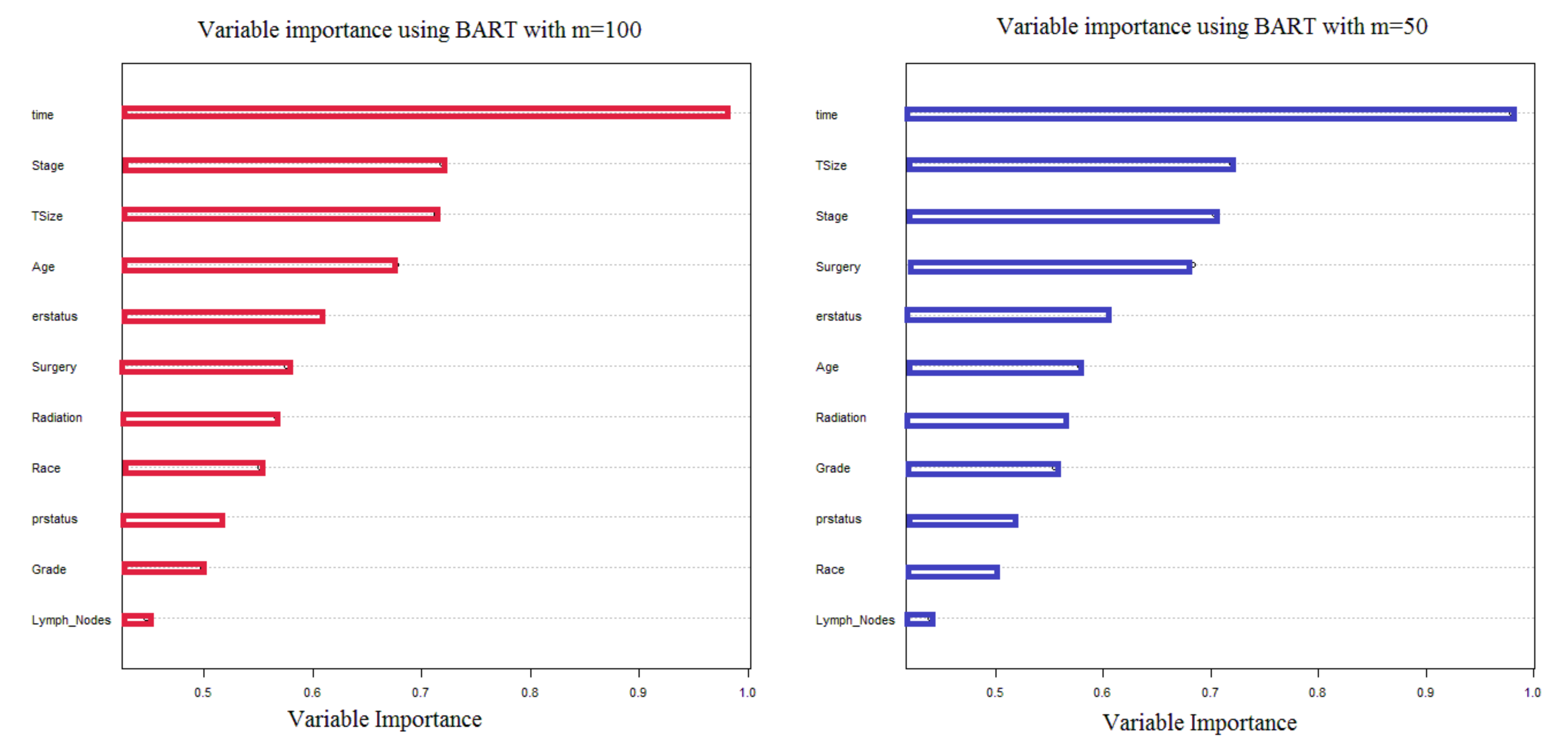}
\caption{Variable importance using BART with 100 and 50 trees.}
\label{VimpBart}
\end{figure}

Finally, we carried out variable selection as described in \ref{method:variables} by examining the average frequency per splitting rule for all 11 predictor variables (time post diagnosis plus 10 predictors), the model being run using different numbers of trees ($m=100,50,20$). As can be seen from Figure \ref{VimpBart},  covariate time naturally is the most selected covariate across the different number of trees. Besides time, the model identifies stage, tumor size, age, ER status and surgery as the five most important covariates impacting overall survival.

We repeated the cross-validation   procedure 20 times, splitting the dataset randomly
into training and test sets in the ratio 2:1, keeping the same censoring rates between training and test sets. Then we used the training set to build the predictor and applied the predictor on the test set to adjudge the performance of the competing methods.

\begin{figure}[ht!]
    \centering
    $\begin{array}{cc}
         \includegraphics[width=0.45\textwidth,trim=2mm 0 12mm 0,clip]{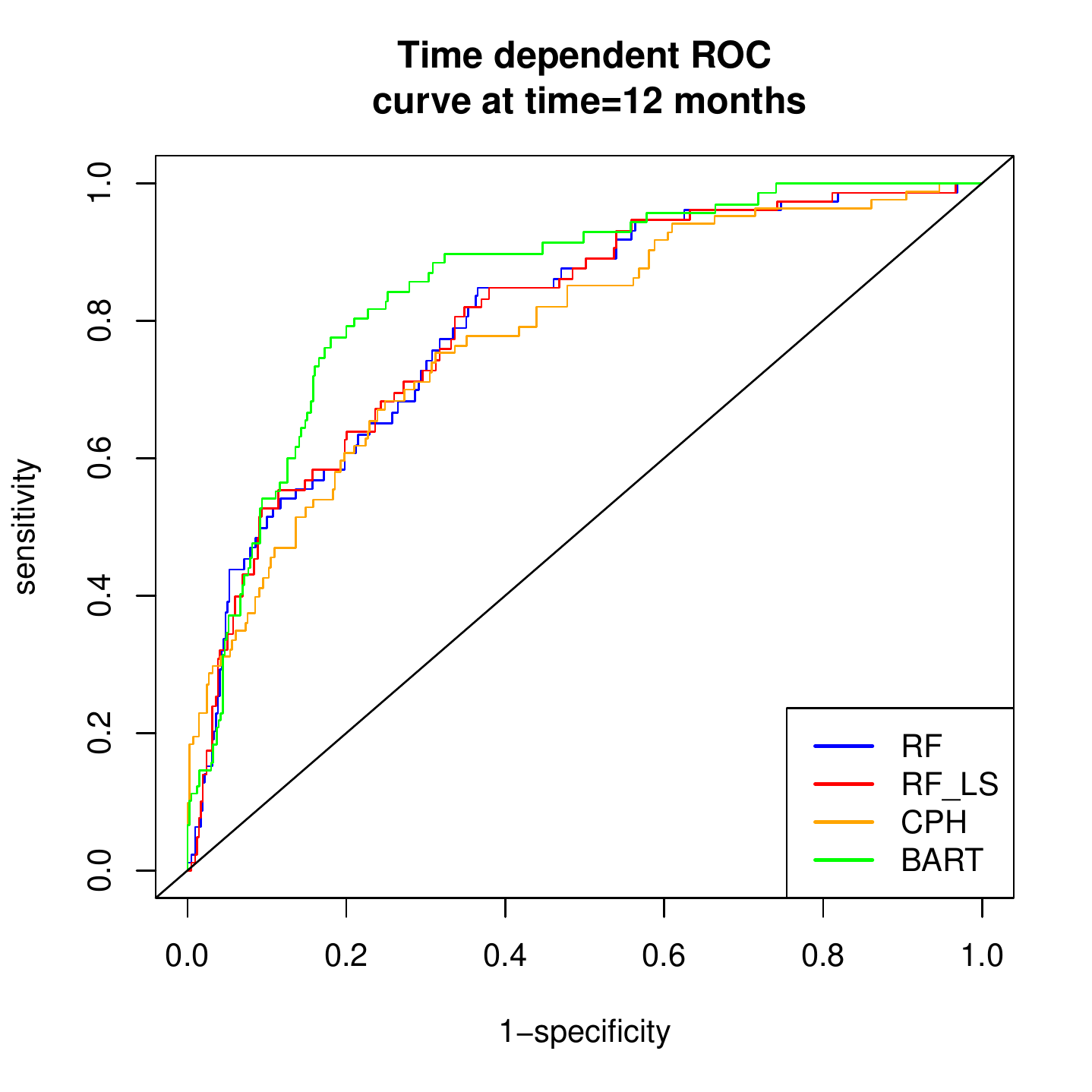} 
         &\includegraphics[width=0.45\textwidth,trim=2mm 0 12mm 0,clip]{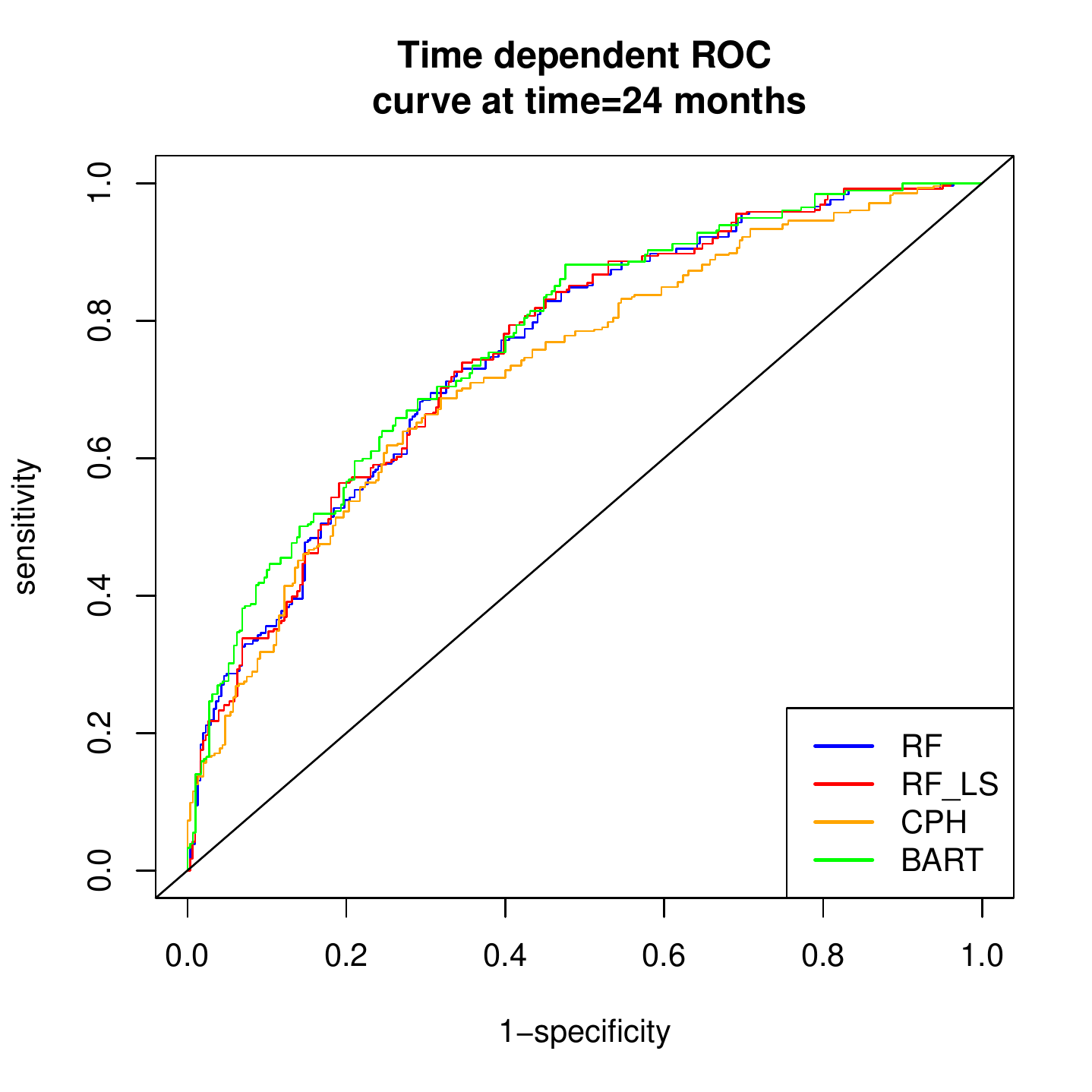} \\
         \includegraphics[width=0.45\textwidth,trim=2mm 0 12mm 0,clip]{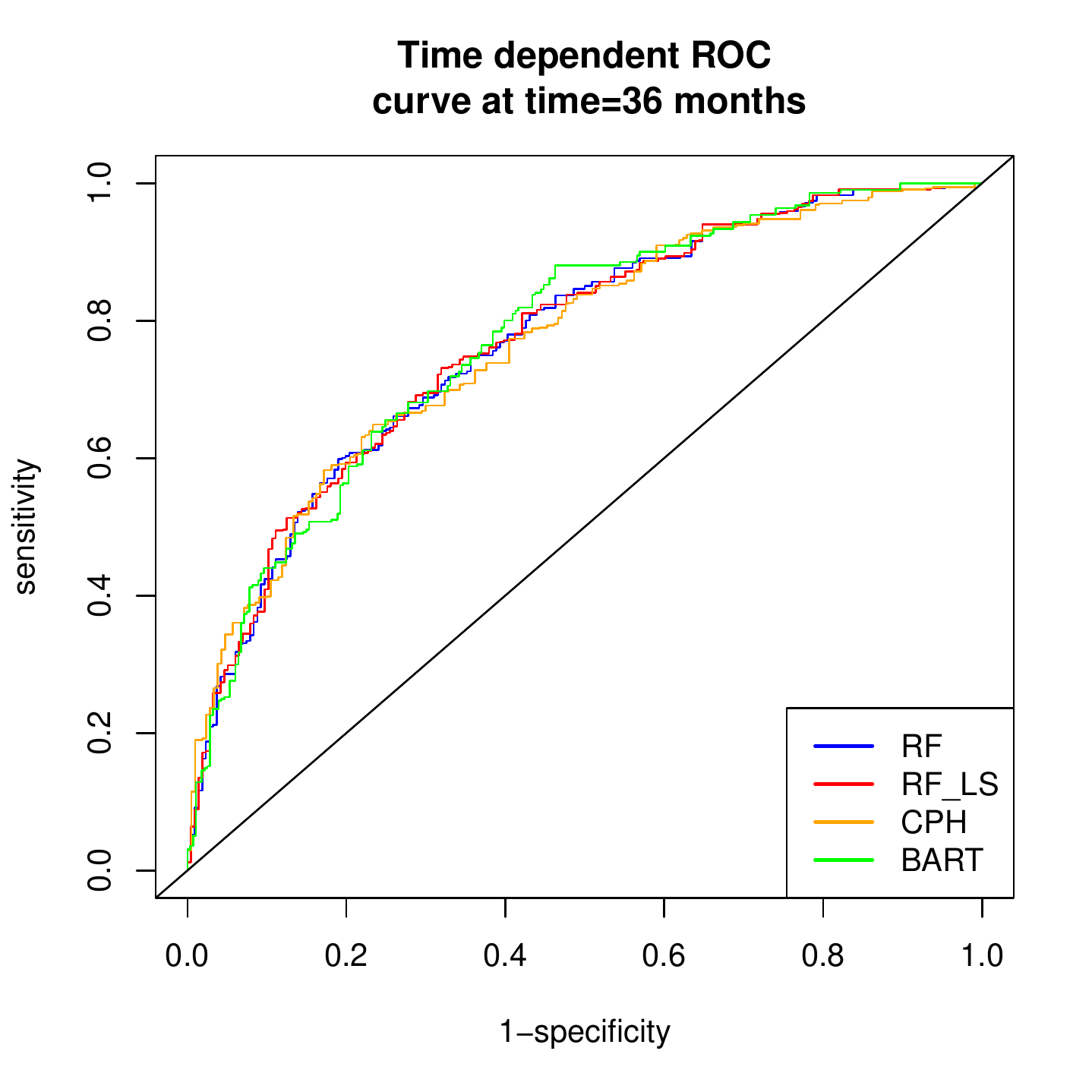} 
         &\includegraphics[width=0.45\textwidth,trim=2mm 0 12mm 0,clip]{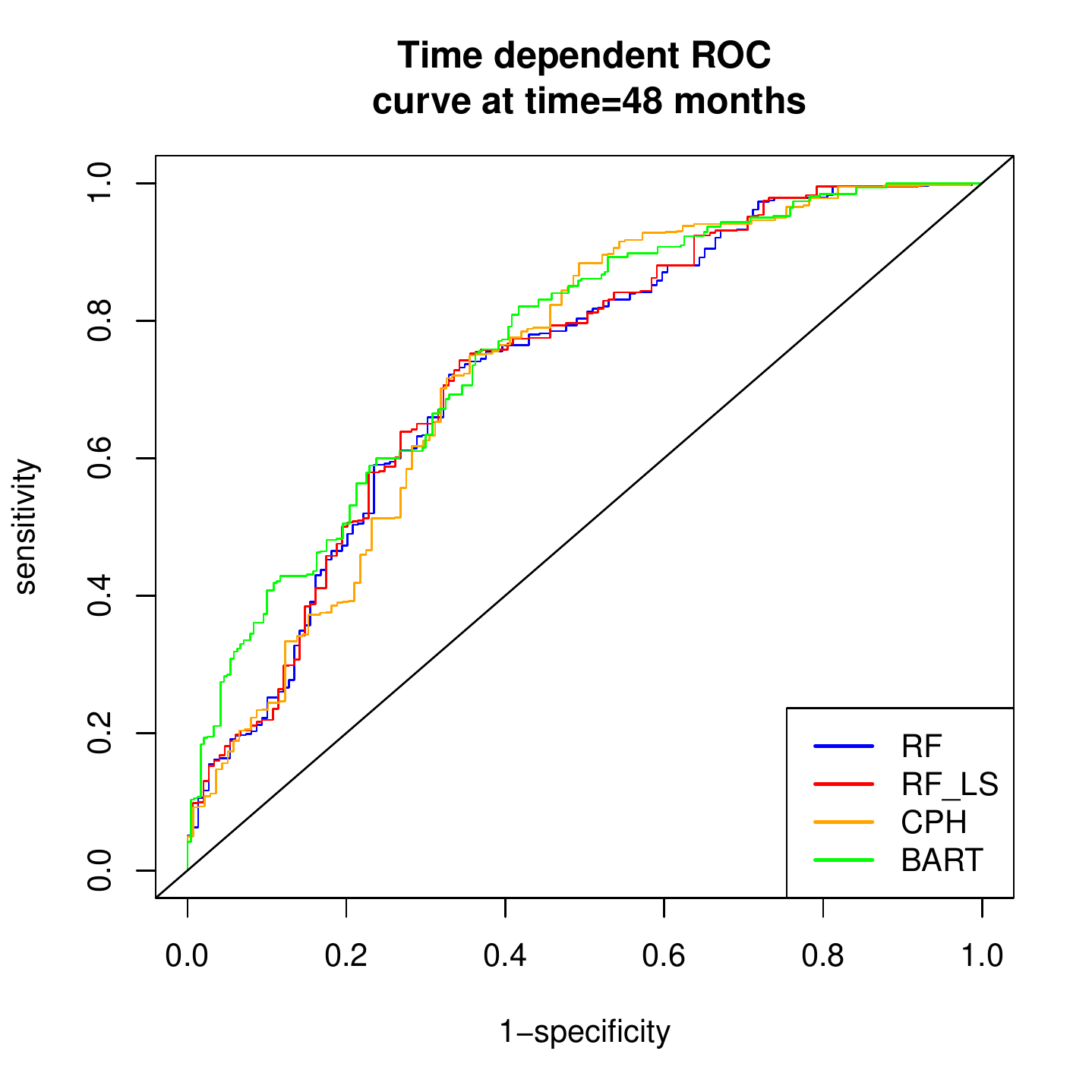} 
    \end{array}$
    \caption{ROC curves over 12, 24, 36 and 48 months}
    \label{AUC plot}
\end{figure}

The performance of the RSF model using log-rank splitting was marginally better as compared to its counterparts (Table \ref{table:PAM}). Values for C-index calculated for all the models indicated that all the estimates were different from 0.5, implying greater capacity of predicting higher probabilities of survival for higher observed survival times. From the plotted time dependent ROC curves in Figure \ref{AUC plot}, we can see that over the first 12 and 24 months of follow up, the BART model has the highest AUC. At time=36 and time=48 months the RSF model with log-rank splitting has the highest AUC.
The estimated AUC value for the RSF model with log-rank splitting tends to decline over time to 0.785 for $12 < t \leq 24$ (Table \ref{table:AUC values}). Thus the estimated AUC values suggests good short-term discriminatory potential of the model score.  Estimates of AUC(t) also become increasingly variable over time due to the diminishing size of the risk set.  Using a follow up of 106 months yields a IAUC estimate of  0.839 for the RSF model with log-rank splitting (Table \ref{table:PAM}). 
 \begin{center}
\begin{table}[ht!]
\centering
\caption{Performance for AUC for the test set}
\label{table:AUC values}
\vspace{4mm}
\begin{tabular}{l@{\hskip 0.5in}cccc}
 \hline 
 & \multicolumn{4}{c}{Time (months)}   \\
\cline{2-5}
Model& 12 & 24 & 36 &  48  \\
\hline
CPH & 0.766 & 0.773 & 0.806 & 0.810\\
 \hline
 BART & 0.839 & 0.789 & 0.811 &0.813\\
  \hline
 RF & 0.807 & 0.785 & 0.821 &0.822\\
   \hline
 RF-LS & 0.782 & 0.771 & 0.813 &0.816 \\
   \hline
    \end{tabular}
 \end{table}
\end{center}

This implies that conditional on one event occurring within 106 months, the probability that the model score is larger for the subject with the smaller event time is 83.9\%. The integrated Brier score values  between 0 and 106 months, for the test set, are lowest for random survival forest with log-rank score splitting. The CPH, BART and RSF  with log-rank splitting models have approximately the same IBS values (Table \ref{table:PAM}).

\begin{center}
\begin{table}
\caption{ Performance assessment for risk prediction for each of the train and test set}
 \label{table:PAM}
\centering
\vspace{4mm}
\begin{tabular}{l@{\hskip 0.5in}cccccccc}
\hline
 &\multicolumn{2}{c}{C-index} && \multicolumn{2}{c}{IAUC} && \multicolumn{2}{c}{IBS}\\
\cline{2-3}\cline{5-6}\cline{8-9}
 Model     & Train & Test &&   Train & Test &&   Train & Test \\ 
 \hline 
CPH & 0.730 &0.722 && 0.850 & 0.839 && 0.112 & 0.119 \\
 \hline
 BART & 0.761 & 0.702 && 0.895 & 0.876 && 0.063 & 0.115 \\
  \hline
 RF & 0.856 & 0.731 && 0.893 & 0.852 && 0.065 & 0.113 \\
   \hline
 RF-LS & 0.825 & 0.726 && 0.889 & 0.849 && 0.065 & 0.112  \\
   \hline
\end{tabular}
\end{table}
\end{center}
All three models perform substantially better than Kaplan-Meier having an IBS estimate of 0.150. Based on all these evaluation measures, it can be inferred that the BART method improves survival prediction accuracy in some cases or has comparable  performance to the other two methods. This improvement could be due to BART's ability to naturally account for additive and non-linear effects.

\section{Conclusion}

\label{sec:conc}This paper focuses on the comparison of BART, CPH and RSF models in analyzing survival data. It reviews  three modeling approaches and compares them in terms of interpretative competence, prediction accuracy and variable selection methods. Simulation studies are performed to judge the competence of the ensemble algorithms with CPH model in regression scenarios. In regression scenarios the three models perform closely when the proportional hazards assumption is met. However the performance of the CPH model depreciates with respect to the other two models when the proportional hazards assumption is violated.

We then apply the three models to analyze a real life breast cancer dataset.The covariates selected for our analysis of the breast cancer survival times fail to follow the proportional hazards assumption of the CPH model. Thus we apply RSF and BART to our dataset to deal with its complex structure and attain increased accuracy in predicting survival times. We chose BART because of its  flexibility to accommodate high dimensional datasets and account for non-linearity and interactions present in covariates. Additionally working under a Bayesian paradigm allowed for natural quantification of uncertainty, that helped in construction of credible and prediction intervals. Thus, regressing on selected predictors we could estimate the median survival time and credible intervals, for a given patient, using the posterior distributions of the process parameters obtained using the BART model. Alternatively survival curves along with confidence bounds for the population could be plotted using all or a subset of covariates. The RSF model was chosen as a competing method to the BART model because similar to BART, it is a decision tree structured black box model having high prediction accuracy and an efficient variable selection mechanism. Being black box models RSF and BART lack interpretative capacities. It cannot directly quantify the risks presented by individual covariates to the overall hazard like the CPH model does in terms of hazard ratios. However the partial dependence survival functions do give us an idea about how each of the covariates individually and jointly affect the overall survival risk. Additionally the performance of all the three models were compared using several assessment measures. BART's and RSF's comparable values of C-index, IAUC and IBSC further validates BART's predictive ability. BART's lack of interpretability as compared to the CPH model can thus be counterbalanced by its gains in prediction accuracy and the ability to incorporate complex interaction effects among the covariates.

Our primary motivation in using the breast cancer dataset was that we were more interested in identifying a statistical model that predicts overall survival effectively based on a set of covariates. We also
wanted to understand the impact of these clinical covariates on the survival of breast cancer patients; and that was carried out successfully by the variable selection methods of BART. 
 We discovered important associations between stage of the tumor, tumor size, er status, surgery status and long-term survival using the BART model. The RSF model additionally considered tumor grade as important. Age at diagnosis was considered an important predictor by the CPH  model. There are many studies which have estimated the risk factor importance of breast cancer using CPH and machine learning models. In line with our findings for the CPH model, \cite{rosenberg2005effect} concluded that tumor size, tumor grade and race,  all have significant constant effects on disease-specific survival in breast cancer, while the effects of age at diagnosis and disease stage have significant effects that do not follow the proportional hazards assumption. Also similar to our results for the RSF and BART models \cite{d2001prognostic} reported tumor size and tumor grade as the most informative medical factors using a RSF model and \cite{delen2005predicting} affirmed the effectiveness of tumor size and tumor stage through the sensitivity analysis of Artificial Neural Network. In contrast with our findings \cite{omurlu2009comparisons} reported the importance of pr status and number of lymph nodes using a CPH and RSF model for analysis.

One of the serious disadvantages of BART in comparison to RSF was its computational time. BART is highly computationally demanding because the model requires  expanding the data at a grid of event times. This problem is aggravated in case of large datasets. The authors mention using parallel processing and time scale coarsening as possible remedies. Both the BART and RSF models have been incorporated as R packages survbart and randomForestSRC respectively and the function computing the RSF model algorithm is  a lot faster.









\bibliographystyle{asa}
\bibliography{library}
\end{document}